\documentclass[conference]{IEEEtran}
\usepackage{amsmath, amssymb, bm, cite, epsfig, psfrag}
\usepackage{graphicx}
\usepackage{soul} %didn't have package _ojas
\usepackage[margin=0.7in]{geometry}
\usepackage{dblfloatfix}
\usepackage{array}
\usepackage{lipsum}%didn't have package _ojas
\newcolumntype{P}[1]{>{\centering\hspace{0pt}}p{#1}}
\newcolumntype{M}[1]{>{\centering\hspace{0pt}}m{#1}}
\newcolumntype{L}{>{\centering\arraybackslash}m{3cm}}
\usepackage[font=small]{caption}
\usepackage{epstopdf}	
\usepackage{longtable}
\usepackage{supertabular,booktabs}
\usepackage{bbm}%didn't have package _ojas
\usepackage{multirow}
\usepackage[usenames,dvipsnames]{xcolor}

\usepackage{subfigure}%didn't have package _ojas
\usepackage{etoolbox}
\usepackage{pbox}
\usepackage{fixltx2e}
\usepackage{tabu}	%didn't have package _ojas
\usepackage{filecontents}
\usepackage{enumerate}
\usepackage{textcomp}
\usepackage{colortbl}
\usepackage{fancyhdr}

\usepackage{bm}
\newtoggle{conference}
\togglefalse{conference} % Use for arxiv version -- also change documentclass
\graphicspath{{figures/}}

\fancypagestyle{firststyle}{
	\fancyhf{}
	\fancyhead[L]{Y. Xing, O. Kanhere, S. Ju, and T. S. Rappaport, "Indoor Wireless Channel Properties at Millimeter Wave and Sub-Terahertz Frequencies," 2019 \textit{IEEE Global Communications Conference (GLOBECOM)}, Hawaii, USA, Dec. 2019, pp. 1-6.}     %% C or L or R.
}
\IEEEoverridecommandlockouts

\begin{document}
\title{Indoor Wireless Channel Properties at Millimeter Wave and Sub-Terahertz Frequencies }
\author{\IEEEauthorblockN{Yunchou Xing, Ojas Kanhere, Shihao Ju, and Theodore S. Rappaport}%\vspace{-0.7cm}

\IEEEauthorblockA{	\small NYU WIRELESS, NYU Tandon School of Engineering, Brooklyn, NY, 11201\\
				\{ychou, ojask, shao, tsr\}@nyu.edu}
				\vspace{-2.5em}	\thanks{This research is supported by the NYU WIRELESS Industrial Affiliates Program and two National Science Foundation (NSF) Research Grants: 1702967 and 1731290.}
}

\maketitle
\thispagestyle{firststyle}

\begin{abstract} 
This paper provides indoor reflection, scattering, transmission, and large-scale \textcolor{black}{path loss} measurements and models, which describe the main propagation mechanisms at millimeter wave and Terahertz frequencies. Channel properties for common building materials (drywall and clear glass) are carefully studied at 28, 73, and 140 GHz using a wideband sliding correlation based channel sounder system with rotatable narrow-beam horn antennas. Reflection coefficient is shown to linearly increase as the incident angle increases, and lower reflection loss (e.g., stronger reflections) are observed as frequencies increase for a given incident angle. Although backscatter from drywall is present at 28, 73, and 140 GHz, smooth surfaces (like drywall) are shown to be modeled as a simple reflected surface, since the scattered power is 20 dB or more below the reflected power over the measured range of frequency and angles. Partition loss tends to increase with frequency, but the amount of loss is material dependent. Both clear glass and drywall are shown to induce a depolarizing effect, which becomes more prominent as frequency increases. Indoor propagation measurements and large-scale indoor path loss models at 140 GHz are provided, revealing similar path loss exponent and shadow fading as observed at 28 and 73 GHz. The measurements and models in this paper can be used for future wireless system design and other applications within buildings for frequencies above 100 GHz. 
\end{abstract}
    

\section{Introduction}~\label{sec:intro}
The use of 5G millimeter wave (mmWave) in wireless communication provides multi-Gbps data rates and enables various new applications like wireless cognition and positioning \cite{rappaport2013millimeter,rappaport19access}. This year (2019) promises to be the ``First year of the 5G era'' \cite{B5GS19}.

In March 2019, the Federal Communications Commission (FCC) voted to open up spectrum above 95 GHz for the first time ever in the USA to encourage the development of new communication technologies and expedite the deployment of new services (ET Docket No. 18-21 \cite{FCC19a}), and provided 21.2 GHz of spectrum for unlicensed use. This ruling provides a partially harmonized unlicensed band at 120 GHz with Japan \cite{nagatsuma2014breakthroughs,JapanmmWave}.  The Institute of Electrical and Electronics Engineers (IEEE) formed the IEEE 802.15.3d \cite{802.15.3d} task force in 2017 for global Wi-Fi use at frequencies across 252 GHz to 325 GHz, creating the first worldwide wireless communications standard for the 250-350 GHz frequency range, with a nominal PHY data rate of 100 Gbps and channel bandwidths from 2 GHz to 70 GHz \cite{802.15.3d}. The use cases for IEEE 802.15.3d include kiosk downloading \cite{petrov16b}, intra-device radio communication \cite{petrov16a}, connectivity in data centers, and wireless fiber for fronthaul and backhaul \cite{coalition19a,802.15.3d,sengupta2018terahertz}. Meanwhile, FCC will launch its largest 5G spectrum auction on December 10, 2019 with 3400 MHz of spectrum in three different bands--37 GHz, 39 GHz, and 47 GHz \cite{FCC18-180}.

Frequencies from 100 GHz to 3 THz are promising bands for the next generation of wireless communication systems because of the wide swaths of unused and unexplored spectrum. Availability of this new spectrum above 95 GHz will open up much needed broadband service enabling new applications for medical imaging, spectroscopy, new massively broadband IoT, sensing, communications, and “wireless fiber” links in rural areas \cite{Mac17JSACa,coalition19a,rappaport19access}. Early work shows that weather and propagation impairments are not very different from today's mmWave all the way up to 400 GHz \cite{Rap11a, rappaport2013millimeter, rappaport19access}.

At mmWave and  THz frequencies, the wavelength $\lambda$ becomes small, motivating the use of hybrid beamforming for ``practical antenna packaging'' \cite{sun2018hybrid,rappaport19access}. At sub-THz, $\lambda$ is comparable to or smaller than the surface roughness of many objects, which suggests that scattering may not be neglected like it was when compared to reflection and diffraction at microwave frequencies (300 MHz to 3 GHz) \cite{ju19icc}.  

Maximum transmission rates of several tens of Gbps for line of sight (LOS) and several Gbps for non-LOS (NLOS) paths were shown to be achievable at 300-350 GHz \cite{kleine12measurement}. Measurements at 100, 200, 300, and 400 GHz using a 1 GHz RF bandwidth channel sounder showed that both indoor LOS and NLOS links (specular reflection from interior building walls) could provide a data rate of 1 Gbps \cite{ma18channel}. Signals with larger incident angles were shown to experience less loss due to the combined effects of reflection, absorption, and scattering. The scattering loss of bare cinderblock walls at 400 GHz was shown to be negligible \cite{ma18channel}.  

\textcolor{black}{There are also notable differences and challenges seen for frequencies beyond 100 GHz (e.g., high phase noise and Doppler, limited output power, and more directional beams), which makes the propagation more challenging \cite{rappaport19access}. Therefore, channel properties at 28, 73, and 140 GHz are studied and compared in this paper.} 

The rest of this paper is organized as follows: Section \ref{sec:SR} presents reflection and scattering measurements and results at 28, 73, and 142 GHz, which show the variation of electrical parameters with frequencies. Section \ref{PLMea} provides an overview of the previous research on partition loss, and presents free space path loss (FSPL) measurement results and partition loss measurements of glass and drywall at 28, 73, and 142 GHz associated with antenna cross polarization measurements to analyze the polarization effects on partition loss for different materials and various frequencies. Section \ref{sec:140PL} shows indoor propagation measurements and an initial indoor large-scale path loss model at 140 GHz for both LOS and NLOS environments. Section \ref{conclusion} provides concluding remarks.

\begin{figure}
	\centering
	\includegraphics[width=0.4 \textwidth]{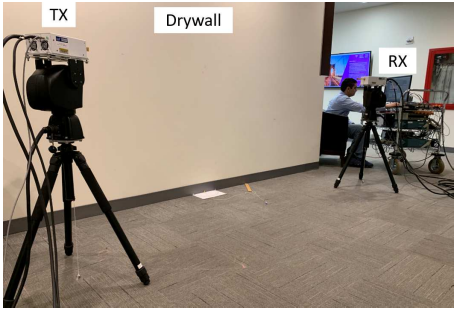}
		\vspace{-1.5 em}
	\caption{Photograph of the reflection/scattering measurement setup. The reflected/scattered power of drywall was measured at a distance 1.5 m away from the wall, in angular increments of  10\textdegree.}
	\label{fig:scatter_meas}
	\vspace{-1.5 em}
\end{figure}

\section{Scattering and Reflection measurements at 28, 73, and 142 GHz}~\label{sec:SR}

\vspace{-1.5em}
\begin{figure*}[htbp]
	\centering
	\subfigure[Measured magnitude of reflection coefficients of drywall at 28 GHz, with $\epsilon_r$ = 4.7 by MMSE estimation.]{
		\begin{minipage}[b]{0.40\textwidth}
			\centering
			\includegraphics[width=1\linewidth]{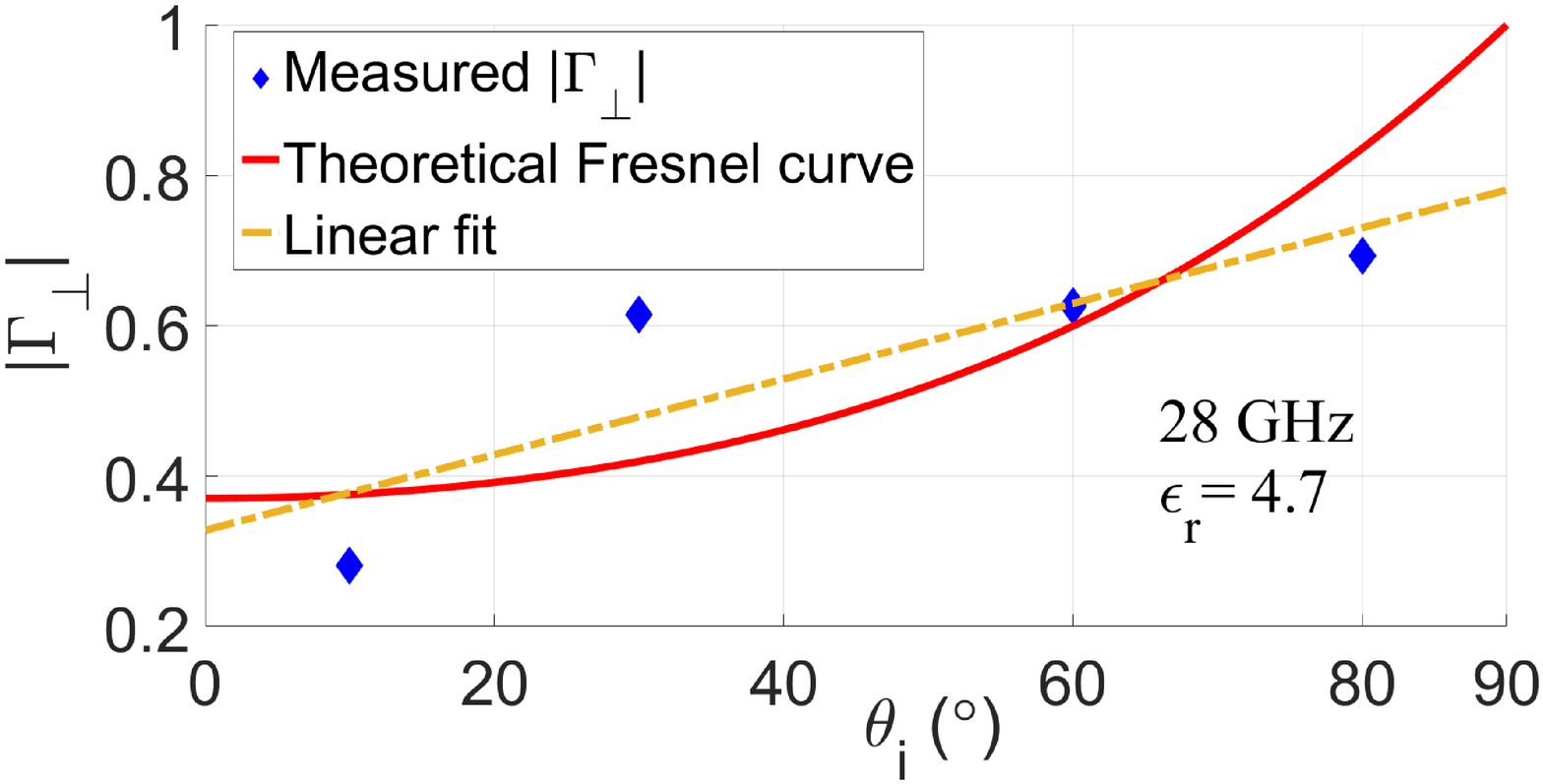}
			\label{fig:RC28}
			\vspace{-1.5em}
		\end{minipage}
	}
	\quad
	\subfigure[Measured reflection/scattering power off drywall at 28 GHz with dual-lobe DS model prediction.]{
		\begin{minipage}[b]{0.40\textwidth}
			\centering
			\includegraphics[width=1\linewidth]{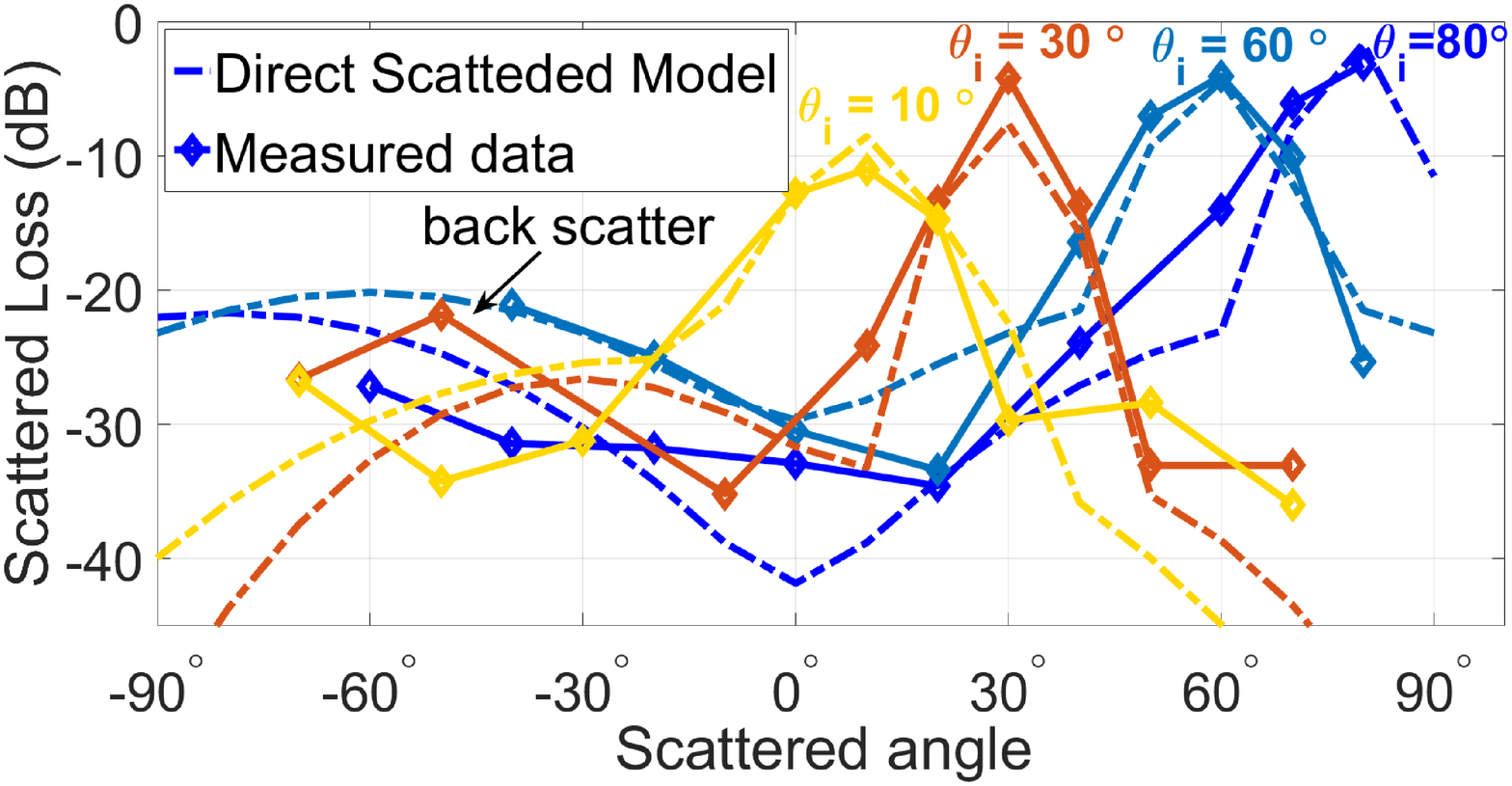}
			\label{fig:SP28}
			\vspace{-1.5em}
		\end{minipage}
	}
	
	%%%%%%%%%%%%%%%%%%%%%%%%%%%%%%%%%%%%%%%%%%%%%%%%%%%%%%%%%%%%%%%%%%%%%%
	\quad
	\subfigure[Measured magnitude of reflection coefficients of drywall at 73 GHz, with $\epsilon_r$ = 5.2 by MMSE estimation.]{
		\begin{minipage}[b]{0.40\textwidth}
			\centering
			\includegraphics[width=1\linewidth]{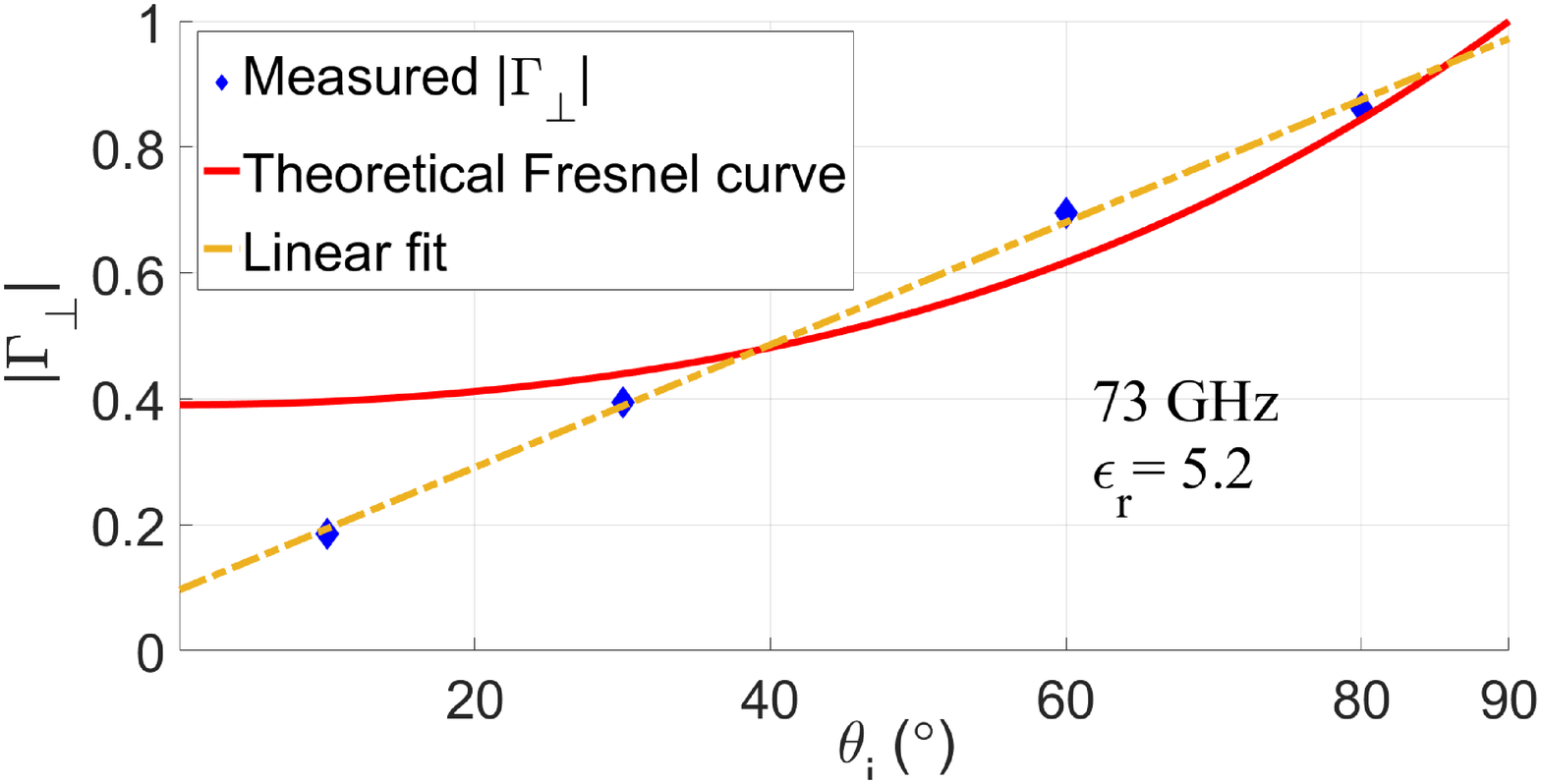}
			\label{fig:RC73}
			\vspace{-1.5em}
		\end{minipage}
	}
	\quad
	\subfigure[Measured reflection/scattering power off drywall at 73 GHz with dual-lobe DS model prediction.]{
		\begin{minipage}[b]{0.40\textwidth}
			\centering
			\includegraphics[width=1\linewidth]{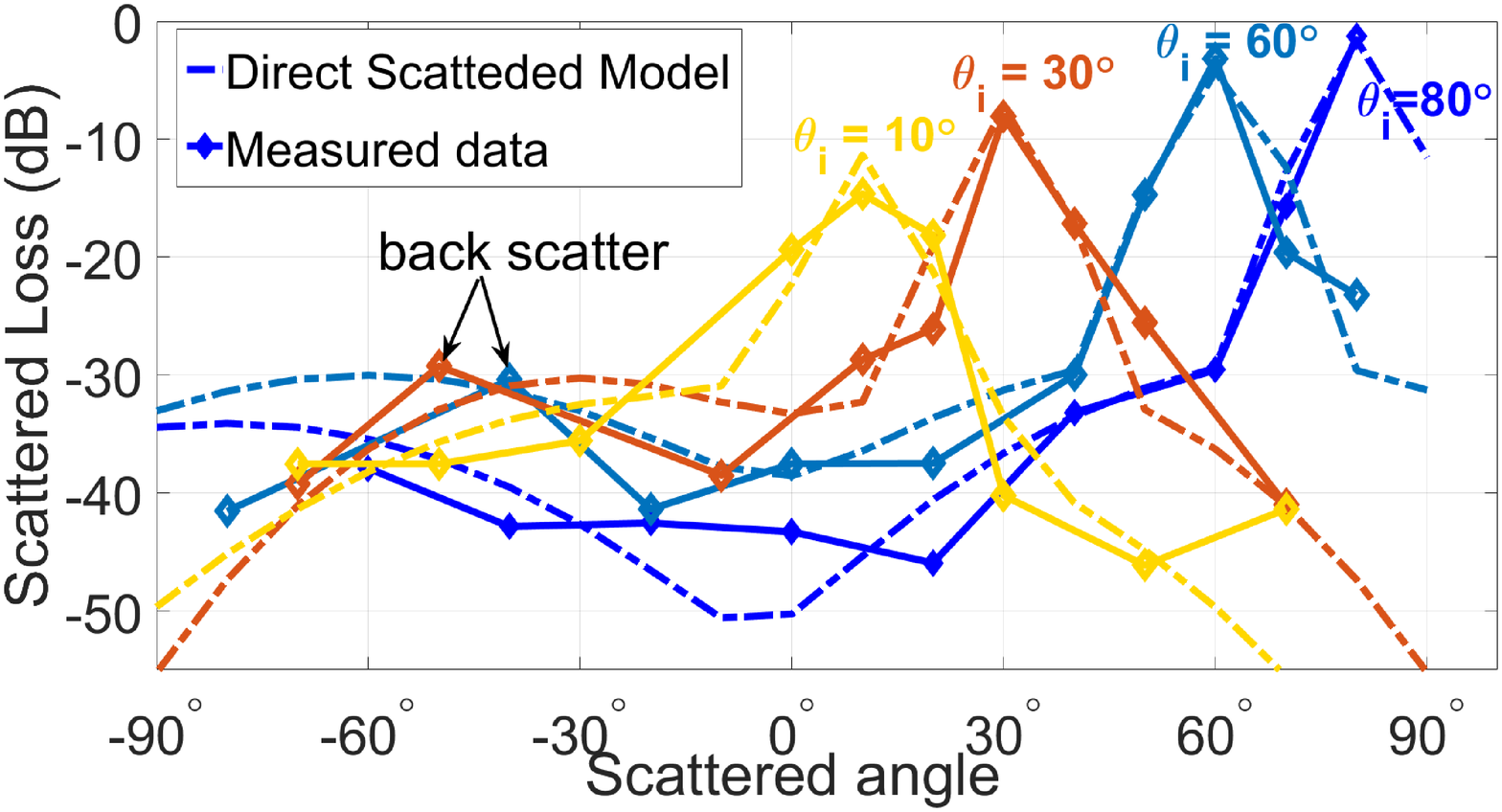}
			\label{fig:SP73}
			\vspace{-1.5em}
		\end{minipage}
	}

	%%%%%%%%%%%%%%%%%%%%%%%%%%%%%%%%%%%%%%%%%%%%%%%%%%%%%%%%%%%%%%%%
	\quad
	\subfigure[Measured magnitude of reflection coefficients of drywall at 142 GHz, with $\epsilon_r$ = 6.4 by MMSE estimation.]{
		\begin{minipage}[b]{0.40\textwidth}
			\centering
			\includegraphics[width=1\linewidth]{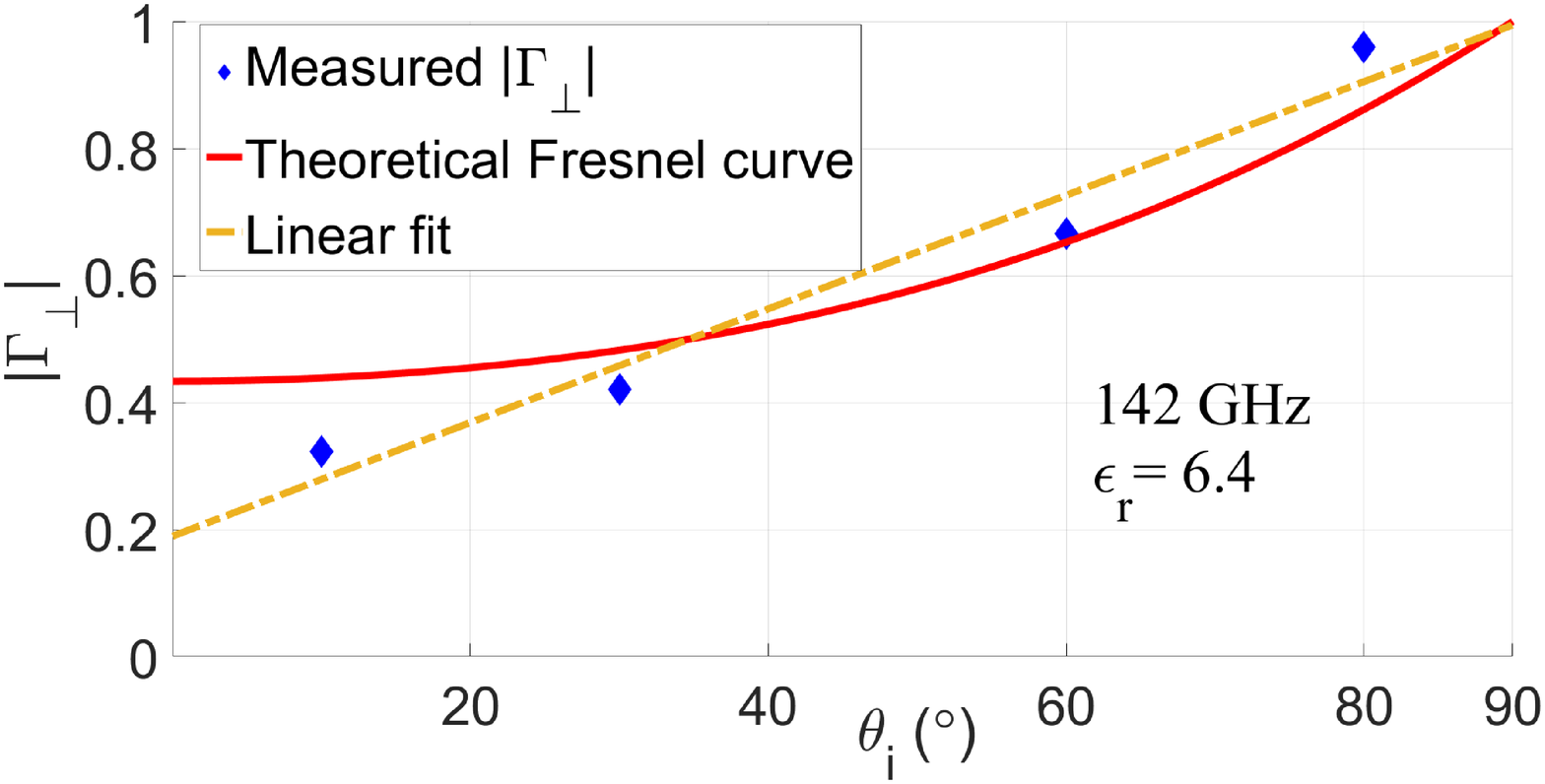}	
			\label{fig:RC140}
			\vspace{-1.5em}
		\end{minipage}
	}
	\quad
	\subfigure[Measured reflection/scattering power off drywall at 142 GHz with dual-lobe DS model prediction.]{
		\begin{minipage}[b]{0.40\textwidth}
			\centering
			\includegraphics[width=\linewidth]{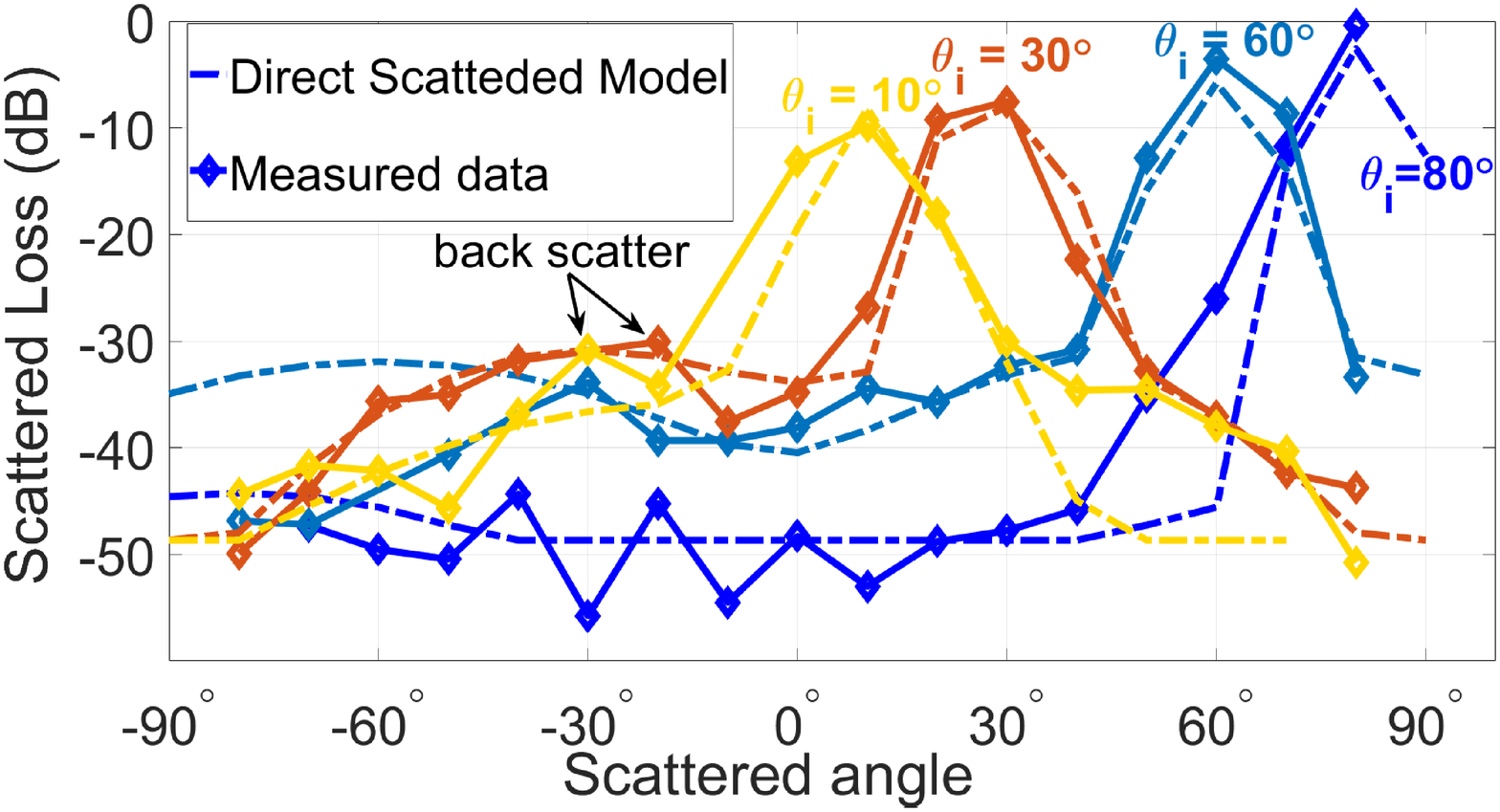}	
			\label{fig:SP140}
			\vspace{-1.5em}
		\end{minipage}
	}
	\quad
	\caption{Comparison between measurements and dual-lobe directive scattering (DS) model plus reflected power using Eq. (3)-(5) and (23) in \cite{ju19icc} at incident angles 10\textdegree, 30\textdegree, 60\textdegree, and 80\textdegree~at 142 GHz for drywall ($\epsilon_r $ = 4.7, 5.2, and 6.4 for drywall  at 28, 73, and 142 GHz. }
	\label{fig:RCSP}
	\vspace{-1.5em}
\end{figure*}

A key to all measurements is using a standard approach for calibration, that assures repeatable measurements by any research team at any frequency \cite{xing18VTC}. 142 GHz FSPL verification measurements were conducted at transmitter-to-receiver (TR) separation distances of 1, 2, 3, 4, and 5 m using the standard calibration and verification method taught in \cite{xing18VTC}, and the results after subtracting out antenna gains were shown  in Fig. 4 of \cite{xing18GC}. The architectures of the channel sounder system used in the measurements given here have been extensively described in \cite{xing18GC,Mac17JSACb, rappaport2013millimeter} and the specifications of the channel sounder system are summarized in Table \ref{tab:1}. The measured path loss at 142 GHz agrees well with Friis FSPL equation \cite{friis1946note}, indicating the high accuracy and proper calibration of the channel sounder system. The close-in (CI) path loss model with 1 m reference distance \cite{rappaport2013millimeter} fits perfectly to the measured data, indicating that the CI model is viable well above 100 GHz.

\subsection{Measurements Setup}
One of the earliest studies of differences between microwave and mmWave frequencies (1.7 GHz vs. 60 GHz) was presented in \cite{davies91a}, which showed the variation of the electrical parameters (e.g., reflection coefficient, conductivity, etc.) of the building materials with frequencies \textcolor{black}{and temperature}.

Since little is known about scattering at mmWave and THz frequencies, reflection and scattering measurements of drywall at 28, 73, and 142 GHz were conducted and the measurement setup is shown in Fig. \ref{fig:scatter_meas}. 
During the measurements, both the heights of TX and RX were set at 1.2 m on an arc with a radius of 1.5 m (which is greater than the Fraunhofer distance) to ensure the propagation is happened in far field \cite{xing18VTC}. Narrow beam horn antennas with HPBWs of 10\textdegree, 7\textdegree, and 8\textdegree, which help to provide high angular resolution, were used at both the TX and RX at 28, 73, and 142 GHz, respectively. Incident angles of $\theta_i$ = 10\textdegree, 30\textdegree, 60\textdegree, and 80\textdegree~were chosen to measure the reflected and scattered power off  drywall, with the incident angle varying from a small angle to a large angle with respect to a line normal to the wall \cite{ju19icc}. The received power was measured from 10\textdegree~to 170\textdegree~(the received power at 0\textdegree~and 180\textdegree~are not able to measure due to the physical size of the antenna). 

\begin{table}[]\caption{Summary of channel sounder systems and antennas used in measurements at 28 GHz, 73 GHz and 142 GHz \cite{rappaport2013millimeter,Mac17JSACb,xing18GC}}\label{tab:1}
	\centering
	\begin{tabular}{llllll}
		\hline
		\multicolumn{1}{|l|}{RF Frequency}   & \multicolumn{1}{l|}{ RF BW} &  \multicolumn{1}{l|}{Antenna} & \multicolumn{1}{c|}{Antenna Gain} & \multicolumn{1}{c|}{ XPD}\\ 
		\multicolumn{1}{|c|}{(GHz)}   & \multicolumn{1}{c|}{ (GHz)} &  \multicolumn{1}{c|}{HPBW} & \multicolumn{1}{c|}{(dBi)} & \multicolumn{1}{c|}{ (dB)}\\ \hline
		\multicolumn{1}{|c|}{28  \cite{rappaport2013millimeter}}           & \multicolumn{1}{c|}{1 }       & \multicolumn{1}{c|}{30\textdegree~/ 10\textdegree}           & \multicolumn{1}{c|}{15.0  / 24.5 }       & \multicolumn{1}{l|}{19.30 }       \\ \hline
		\multicolumn{1}{|c|}{73  \cite{Mac17JSACb}}      & \multicolumn{1}{c|}{1 }     & \multicolumn{1}{c|}{15\textdegree~/ 7\textdegree}           & \multicolumn{1}{c|}{20.0  / 27.0 }       & \multicolumn{1}{c|}{28.94 }       \\ \hline
		\multicolumn{1}{|c|}{142  \cite{xing18GC}}         & \multicolumn{1}{c|}{1 }      & \multicolumn{1}{c|}{8\textdegree}            & \multicolumn{1}{c|}{27.0 }       & \multicolumn{1}{l|}{44.18 }       \\ \hline
	\end{tabular}
\end{table}

\subsection{Reflection at mmWave and THz}
\label{sec:RF}
The Fresnel reflection coefficient $\Gamma_{\perp}$ (when the E-field is normal to the plane of incidence) is given by \cite{Rap02a}:
\begin{equation}~\label{equ:gamma}
\small
\Gamma_{\perp} = \dfrac{E_r}{E_i} = \dfrac{\cos\theta_i-\sqrt{\epsilon_r-\sin^2\theta_i}}{\cos\theta_i+\sqrt{\epsilon_r-\sin^2\theta_i}},
\end{equation}
where $E_r$ and $E_i$ are the electric fields of the reflected and incident waves with units of V/m, $\epsilon_r$ is the permittivity of the reflecting surface, and the incident angle $\theta_i$ is defined as the angle between the incident direction and normal. 

Based on the measured data shown in Table \ref{tab:RL} and Fresnel's equation \eqref{equ:gamma}, $\epsilon_r$ = 4.7, 5.2, and 6.4 was obtained through a minimum mean square error (MMSE) estimator of $|\Gamma_{\perp}|^2$ at 28, 73, and 142 GHz, respectively.  As shown in Fig. \ref{fig:RC28}, \ref{fig:RC73}, and \ref{fig:RC140}, the blue diamond points indicate the magnitude of the measured reflection coefficients $|\Gamma_{\perp}|$ and the red lines are the theoretical Fresnel curve through MMSE estimation. It is worth noting that, a linear fit (yellow dashed lines) of the magnitude of reflection coefficient with the incident angle in degrees performs better than the Fresnel equation at these three frequencies. Table \ref{tab:RL} shows that the reflection loss at 142 GHz ranges from 0.36 dB, when the incident angle is close to grazing ($\theta_i$= 80\textdegree), to 9.81 dB when the incident direction is nearly perpendicular to the surface of drywall ($\theta_i$= 10\textdegree), and the reflection loss linearly decreases as the incident angles $\theta_i$ increases. It is observed that reflections are stronger at higher frequencies (the permittivity $\epsilon_r$ is smaller at lower frequencies).     % Maybe add a table of reflection loss

\begin{table}
	\centering
	\caption{Reflection Loss vs. Frequencies \& Angles}\label{tab:RL}
	\begin{tabular}{|c|c|c|c|c|} 
		\hline
		f / $\theta_i$& 10\textdegree     & 30\textdegree    & 60\textdegree    & 80\textdegree     \\ 
		\hline
		28 GHz  & -12.98 dB& -4.22 dB& -4.06 dB & -3.18 dB \\ 
		\hline
		73 GHz & -12.65 dB & -8.08 dB& -3.16 dB& -1.28 dB \\ 
		\hline
		142 GHz& -9.81 dB& -7.53 dB& -3.54 dB& -0.36 dB\\
		\hline
	\end{tabular}
	\vspace{-1.5em}
\end{table}

\subsection{Scattering at mmWave and THz}
Measured scattering patterns of different incident angles at 28, 73, and 142 GHz are shown in Fig. \ref{fig:SP28}, \ref{fig:SP73}, and \ref{fig:SP140}, respectively. The peak measured power (scattered power plus reflected power) was observed to occur at the specular reflection angle. The peak measured power was greater at larger incident angles than at smaller incident angles (e.g., 9.4 dB difference between 80\textdegree~and 10\textdegree at 142 GHz), where most of the energy is due to reflection but not scattering \cite{rappaport19access}. At all angles of incidence, measured power was within 10 dB below the peak power in a $ \pm$ 10\textdegree~angle range of the specular reflection angle, likely a function of antenna patterns. In addition, backscattered power was observed (e.g., 10\textdegree~and 30\textdegree~incidence at 142 GHz) but was more than 20 dB below the peak received power, which means that the surface of drywall can still be consider to be smooth even at 142 GHz and the specular reflection is the main mechanism for indoor propagation at 142 GHz. 

Comparisons between measurements and predictions made by a dual-lobe directive scattering (DS) model (as introduced in \cite{ju19icc,Esposti07a}) with TX incident angle $\theta_i$= 10\textdegree, 30\textdegree, 60\textdegree, and 80\textdegree~are shown in Fig. \ref{fig:SP28}, \ref{fig:SP73}, and \ref{fig:SP140}. Permittivity $\epsilon_r$ = 4.7, 5.2, and 6.4 estimated from the reflection measurements using \eqref{equ:gamma}, are used in the dual-lobe DS model at 28, 73, and 142 GHz, respectively. It can be seen that simulations of peak received power (the sum of reflection and scattering) at the specular reflection angle agrees well with measured data (within 3 dB), confirming that scattering can be modeled approximately by a smooth reflector with some loss  (see (3)-(5) and (23) in \cite{ju19icc}) when material properties are known, while scattering at other scattering angles falls off rapidly.

\section{Partition Loss at 28, 73, and 142 GHz}~\label{PLMea}
In addition to reflection and scattering, transmission/penetration is another important mechanism for wireless communication systems at mmWave and THz frequencies. Partition loss is defined as the difference between signal power right before the partition and the signal power right after the partition \cite{anderson04a}, which includes reflection/scattering loss and the material absorption loss. The partition loss describes the how the signal power changes after a partition in the radio link. Wideband mmWave and Terahertz networks, as well as precise ray-tracer algorithms, will require accurate channel models that predict the partition loss induced by common building objects\cite{xing18VTC,ojas2018globecomm,Kanhere19a}. Therefore, partition loss of common building materials needs to be extensively investigated for 5G mmWave wireless systems and future Terahertz wireless communications in and around buildings.

\subsection{Antenna XPD Measurements}

\begin{figure}    
	\centering
	\includegraphics[width=0.45\textwidth]{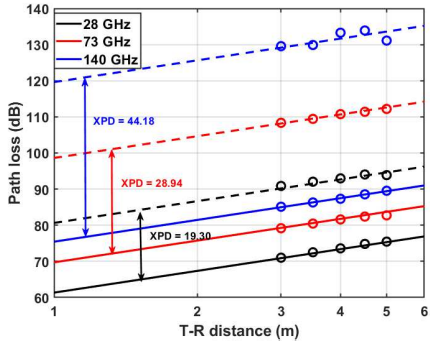}
		\vspace{-2.0 em}
	\caption{Measured antenna XPD at 28, 73, and 142 GHz. The solid lines and the dash lines represent the path loss measured with co-polarized and cross-polarized antennas, respectively. The XPD values calculated across five distances are within 1 dB at each frequency, which validate the XPD measurements.}
	\label{fig:XPD}
		\vspace{-1.0 em}
\end{figure}

In order to measure the partition loss of a material for different polarizations, the antenna cross polarization discrimination (XPD) at different frequencies was measured to analyze the electrical properties of the antennas \cite{xing18VTC,xing18GC}. XPD values also are required to analyze the polarization effects of partitions at different frequencies. 

The XPD measurements were conducted at 28, 73, and 142 GHz in LOS free space first with T-R separation distances in the far-field (e.g., 3-5 m were chosen in this paper) while ensuring the TX and RX antennas are perfectly boresight aligned. There were no nearby reflectors or obstructions present in the propagation path that might cause multipath reflections or induce fading during the measurements, and the heights of the antennas and the T-R separation distances between the antennas were selected to ensure ground bounces and ceiling bounces do not induce reflection, scattering, or diffraction within or just outside the HPBW of the main lobe of the TX/RX antenna \cite{xing18VTC}. After free space power measurements with co-polarized antennas, measurements were then conducted at the same distances but with cross-polarized antennas (e.g., V-H and H-V). Cross-polarization was realized by using a waveguide twist which rotates the antenna by 90\textdegree. The insertion loss caused by the twist was measured and calibrated out. The detailed measurement guidelines and procedures were presented in \cite{xing18VTC}. 

 The path losses using co-polarized and cross-polarized antennas at different frequencies are shown in Fig. \ref{fig:XPD}. The XPD was calculated by taking the difference between the path losses between the co-polarized and cross-polarized antenna configurations at a given distance, as shown in Table \ref{tab:1}. Note that, at a fixed TR separation distance, the free space received powers for the H-H and H-V configurations were within 1 dB of the V-V and V-H received powers, respectively, showing reciprocity with cross-polarization measurements.

\begin{table}
	\centering
	\caption{Partition Loss at 28, 73, and 142 GHz for Clear Glass}~\label{tab:partitionG}
	\footnotesize
	\begin{tabular}{|c|c|c|c|c|c|c|} 
		\hline
		\multirow{3}{*}{Pol.} & \multicolumn{2}{c|}{28 GHz} & \multicolumn{2}{c|}{73 GHz} & \multicolumn{2}{c|}{142 GHz}  \\ 
		\cline{2-7}
		& Mean & STD                      & Mean & STD                          & Mean & STD                            \\ 
				& (dB) & (dB)                      & (dB) & (dB)                          & (dB) & (dB)                            \\ 
		\hline
		V-V                                        & 1.53      & 0.60                              & 7.17      & 0.17                              & 10.22     & 0.22                                \\ 
		\hline
		V-H                                        & 20.63     & 1.32                              & 37.65     & 0.53                              & 46.92     & 2.05                                \\ 
		\hline
		H-V                                        & 22.25     & 0.88                              & 36.92     & 1.11                              & 37.37     & 1.79                                \\ 
		\hline
		H-H                                        & 1.48      & 0.54                              & 7.15      & 0.44                              & 10.43     & 0.55                                \\
		\hline
	\end{tabular}
\vspace{-0.5em}
\end{table}

\begin{table}
	\centering
	\caption{Partition Loss at 28, 73, and 142 GHz for Drywall}~\label{tab:partitionD}
	\begin{tabular}{|c|c|c|c|c|c|c|} 
		\hline
		\multirow{3}{*}{Pol.} & \multicolumn{2}{c|}{ 28 GHz} & \multicolumn{2}{c|}{73 GHz} & \multicolumn{2}{c|}{142 GHz}  \\ 
		\cline{2-7}
		& Mean  & STD                          & Mean &STD                         & Mean  & STD                           \\ 
				& (dB) & (dB)                      & (dB) & (dB)                          & (dB) & (dB)                            \\ 
				\hline
		V-V                                        & 4.15      & 0.59                              & 2.57      & 0.61                              & 8.46      & 1.22                                \\ 
		\hline
		V-H                                        & 25.59     & 2.85                              & 24.97     & 0.58                              & 27.28     & 1.77                                \\ 
		\hline
		H-V                                        & 25.81     & 0.65                              & 23.38     & 0.65                              & 26.00     & 1.42                                \\ 
		\hline
		H-H                                        & 3.31      & 1.13                              & 3.17     & 0.68                              & 9.31      & 0.61                                \\
		\hline
	\end{tabular}
	\vspace{-0.5em}
\end{table}

\subsection{Partition Loss Measurements}

Partition loss measurements at 28, 73, and 142 GHz were conducted at T-R separation distances of 3, 3.5, 4, 4.5, and 5 m, and the TX/RX antenna heights 1.6 m were chosen (refer to Fig. 3 in \cite{xing18VTC}). The separation distances ensure the measured material is in the far-field of the TX and a plane wave is incident upon the material under test (MUT). The dimensions of the MUT were large enough to guarantee that the radiating wavefront from the TX antenna is illuminated on the material without exceeding the physical dimensions of the MUT \cite{xing18VTC}. At each distance, 5 measurements were recorded with slightly movement in the order of half a wavelength, taking the average of the power in the first arriving multipath component of the recorded PDPs, to exclude the multipath constructive or destructive effects. 

The power that gets transmitted through the material and reaches the RX on the other side of the MUT was measured for four types of TX-RX antenna orientation pairs: the V-V orientation, the V-H orientation, the H-V orientation (the TX antenna is horizontally polarized while the RX antenna is vertically polarized), and the H-H orientation.
 
Common building construction materials, drywall (with a thickness of 14.5 cm) and clear glass (with a thickness of 0.6 cm), were selected to be the MUT, with measurement results listed in Table \ref{tab:partitionG} and Table \ref{tab:partitionD}, respectively. The partition losses were measured and calculated as:
\begin{equation}\label{equ:LVV}
\begin{split}
\footnotesize
L_{XY}\text{[dB]} = P_{tX}\text{[dBm]}-P_{rY}(d)\text{[dBm]}-\text{FSPL}(d)\text{[dB]},
\end{split}
\end{equation}
where $L_{XY}\text{[dB]}$ is the material partition loss, $X$, and $Y$ can be either V or H, corresponding to vertically polarized or horizontally polarized antenna configuration at the link ends, $P_{rY}(d)$  is the RX received power in dBm at distance $d$ in meters with the MUT between the TX and RX, $P_{tX}\text{[dBm]}$ is the transmitted power from the TX , and $\text{FSPL}(d)\text{[dB]}$ is the free space path loss at distance $d$ \cite{xing18VTC,friis1946note}. 

\begin{figure}  
	\centering
	\includegraphics[width=0.45\textwidth]{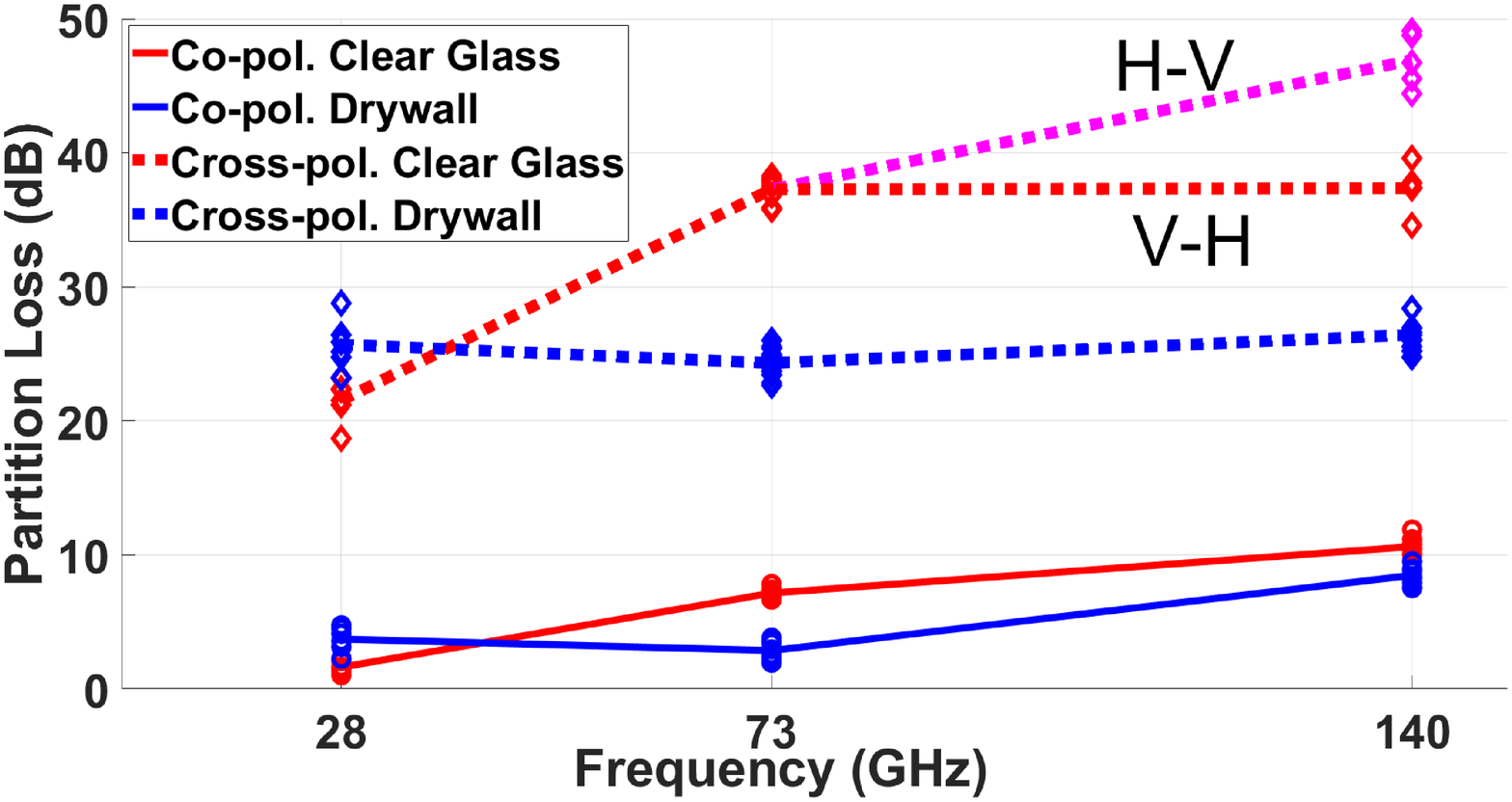}
	\caption{Partition loss measurement results of clear glass with a thickness of 0.6 cm (red lines) and drywall with a thickness of 14.5 cm (blue lines) at 28, 73, and 142 GHz.}
	\label{fig:partition}
		\vspace{-1.5em}  
\end{figure}

The measured mean partition loss of clear glass at 28 GHz for co-polarized situation, see Table \ref{tab:partitionG}, is 1.50 dB with a standard deviation (STD) of 0.50 dB. The mean partition loss for co-polarized situation is 7.16 dB with a STD of 0.15 dB at 73 GHz, and 10.33 dB with a STD of 0.24 dB at 142 GHz. According to the measurements, the partition loss of clear glass increases with the frequencies moderately, as expected, rising from 1.50 dB at 28 GHz to 10.33 dB at 142 GHz. 
 
At 28 and 73 GHz, the difference of clear glass partition losses in cross-polarization situation (V-H and H-V) are negligible. However, at 142 GHz, the mean partition loss of clear glass with V-H configuration is 9.55 dB higher than that with H-V configuration, which means the material has different polarization effects at higher frequencies. It is worth noting that the XPD is not subtracted from the cross-polarized partition loss measurements shown in Table \ref{tab:partitionG} and Table \ref{tab:partitionD}. Subtracting the XPD results in a negative value of partition loss due to the polarization coupling effects (depolarization) of the building materials. Fig. \ref{fig:partition} illustrates that the partition loss of clear glass tends to increase with frequencies for both co- and cross-polarized antenna configurations. 

As shown in Fig. \ref{fig:partition} and Table \ref{tab:partitionD}, the mean partition loss of drywall for co-polarization configuration is 3.73 dB at 28 GHz and 2.87 dB at 73 GHz, respectively. However, the mean partition loss increases to 8.89 dB at 142 GHz. 
 
For cross-polarization configuration, the mean partition losses of drywall at 28, 73 , and 142 GHz are 25.70 dB, 24.18 dB, and 26.64 dB, respectively. There is negligible difference between the partition loss of V-H and H-V configurations. After subtracting the measured XPD values (as shown in Fig. \ref{fig:XPD}), we get 6.40 dB, -4.76 dB, and -17.54 dB, where the negative value means the drywall induce polarization coupling effects (depolarization) at 73 and 142 GHz. 

Work in \cite{ma18channel} showed that absorption imposed a $\sim$8dB penalty to the reflection power (nearly 16\% of the signal power imping on the reflection surface is reflected and about 84\% of the power is absorbed) from an indoor painted cinderblock wall at 100 GHz and the effect of scattering from the painted cinderblock wall is significantly smaller than the effect of absorption. It is worth noting that the absorption mentioned in \cite{ma18channel} includes the power penetrating through the wall and the power absorbed by the wall. 

Using the $\epsilon_r=6.4$ of drywall \textcolor{black}{at 140 GHz} from Section \ref{sec:RF}, a reflection loss of 7.25 dB is predicted ($\sim$18.8\% of the power is reflected), which is comparable to the reflected power ($\sim$8 dB) measured in \cite{ma18channel} at 100 GHz. In addition, as seen in Section \ref{PLMea}, 14.3\% of the incident power is transmitted through drywall at 142 GHz (8.46 dB partition loss was measured). Thus, there is about 66.9\% (100\%-18.8\%-14.3\%) of the power impinging on the surface ($\sim$4.8 dB real absorption loss) was absorbed by drywall at 142 GHz.   

\begin{figure}    
	\centering
	\includegraphics[width=0.45\textwidth]{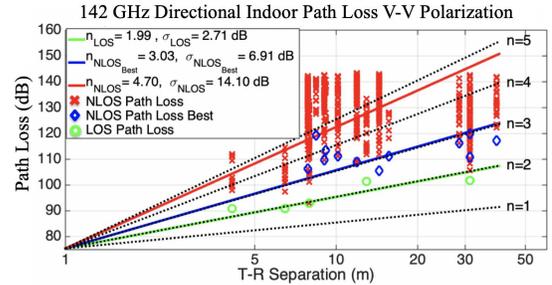}
	\caption{\textcolor{black}{142 GHz directional path loss scatter plot and indoor directional CI ($d_0$= 1 m) path loss model for both LOS and NLOS scenarios. Each green circle represents LOS path loss values, red crosses represent NLOS path loss values measured at arbitrary antenna pointing angles between the TX and RX, and blue diamonds represent angles with the lowest path loss measured for each NLOS TX-RX location combination} \cite{Mac15b}.}
	\label{fig:direc140PL}
		\vspace{-1.5 em}
\end{figure}

\section{Indoor Propagation measurements and Path loss model at 142 GHz}\label{sec:140PL}

Wideband indoor propagation measurements at 142 GHz (see the specification of the 142 GHz channel sounder system in Table \ref{tab:1}) were conducted in a multipath-rich indoor environment at the 9th floor of 2 MetroTech Center using the exact same locations as used at previous 28 and 73 GHz \cite{Mac15b,Deng15a}, which is a typical indoor environment including hallway, meeting rooms, cubical office, laboratory and open area \cite{xing18GC}. The TX antenna were set at 2.5 m near the ceiling (2.7 m) to emulate current indoor wireless access points and the RX antennas were set at heights of 1.5 m which are typical heights of mobile devices. The measurements were conducted with both co-polarized and cross-polarized antennas, and for each TX-RX combination, 3 elevation angles at both TX and RX were chosen (boresight, up tilted by 8\textdegree, and down tilted by 8\textdegree, which cover 95\% of the total power \cite{Sun15a}) and both TX and RX rotated 360\textdegree~in azimuth by 8\textdegree/step to cover the entire azimuth plane \cite{Mac15b}. In the meantime, indoor ray tracing techniques will be used to assist the measurements and will produce simulations together with the measurements to provide an accurate stochastic indoor channel model across different frequencies and various bandwidth \cite{ojas2018globecomm,Kanhere19a}.

Fig. \ref{fig:direc140PL} presents the directional path loss scatter plot and best-fit CI path loss model \cite{rappaport2013millimeter,Mac15b} at 142 GHz for both LOS and NLOS environment. The LOS path loss exponents (PLE) are 1.7 at 28 GHz, 1.6 at 73 GHz, and 2.0 at 142 GHz, as shown in Table \ref{tab:PLcomp}, showing that there is a bit more loss at 142 GHz likely due to atmospheric attenuation \cite{rappaport19access}. The NLOS-Best PLEs and the NLOS PLEs are similar over all three frequencies, respectively, with NLOS at 142 GHz having slightly less loss than lower frequencies, likely due to greater reflected power as frequency increases (see Fig. \ref{fig:RCSP}). Overall, we surmise the 142 GHz indoor path loss models are similar as models at frequencies below 100 GHz. More data will be collected to provide statistical channel impulse response models, as well as outdoor measurements and models above 100 GHz, in the future. 

\begin{table}
	\centering
	\caption{Indoor Directional CI path loss model at 28, 73, and 142 GHz for both LOS and NLOS environment \cite{Mac15b,Deng15a}}~\label{tab:PLcomp}
	\begin{tabular}{|c|c|c|c|c|c|c|} 
		\hline
		\multirow{3}{*}{Env.} & \multicolumn{2}{c|}{ 28 GHz \cite{Mac15b}} & \multicolumn{2}{c|}{73 GHz\cite{Mac15b}} & \multicolumn{2}{c|}{142 GHz}  \\ 
		\cline{2-7}
		& n  & $\sigma$                          & n &$\sigma$                           & n  & $\sigma$                            \\ 
		& (dB) & (dB)                      & (dB) & (dB)                          & (dB) & (dB)                            \\ 
		\hline
		LOS                                        & 1.70      & 2.50                              & 1.60      & 3.20                              & 1.99      & 2.71                                \\ 
		\hline
		$\text{NLOS}_{Best}$                                       & 3.00     & 10.80                              & 3.40     & 11.80                              & 3.03    & 6.91                                \\ 
		\hline
		NLOS                                        & 4.40     & 11.60                              & 5.30     & 15.70                              & 4.70     & 14.10                                \\ 
		\hline
	\end{tabular}
	\vspace{-1.5 em}
\end{table}

\section{Conclusion}\label{conclusion}

This paper investigates reflection and scattering effects with real-world measurements at mmWave and THz frequencies. The reflection loss of indoor drywall is observed to be lower (e.g., reflection are stronger) at higher frequencies and range from 0.4 dB to 9.8 dB at 140 GHz with \textcolor{black}{impinging direction from grazing (e.g., 80\textdegree) to nearly perpendicular to the reflection surface (e.g., 10\textdegree), respectively}. The dual-lobe DS model is shown to provide a good estimation of the scattering power with known electrical parameters of the scattering surface. Backscatter is both modeled and measured to be more than 20 dB down from the peak received power (scattered plus reflected) and to a first order approximation, smooth surfaces like drywall can be modeled as reflective surfaces, especially close to grazing.

Antenna XPD, measured at 28, 73, and 140 GHz, is shown to not change with distance and has a trend to increase with frequencies. The partition loss of clear glass and drywall has been measured at 28, 73, and 140 GHz with four polarization configurations (V-V, V-H, H-V, and H-H), using horn antennas having similar aperture. Measuring the antenna XPD enables the analysis of depolarization effects of clear glass and drywall. Due to signal depolarization, the partition loss for cross-polarized antenna orientations is less than the expected value based on the XPD measurements and the co-polarized partition measurements. The partition loss is highly dependent on antenna polarization for both materials, since both clear glass and drywall induce a depolarizing effect, which becomes more prominent as frequency increases. Ongoing propagation measurements and initial large-scale path loss results show that there is not much difference in the path loss over 28, 73, and 140 GHz, with slight more large-scale LOS path loss at 140 GHz likely due to absorption and slightly less loss over distance in NLOS due to stronger reflections. 

%\section{Acknowledgments}
%This research is supported by the NYU WIRELESS Industrial Affiliates Program and two National Science Foundation (NSF) Research Grants: 1702967 and 1731290.

\bibliographystyle{IEEEtran}
\bibliography{partition_loss5}

\end{document}